\documentclass[twocolumn,secnumarabic,amssymb, nobibnotes, aps, prd]{revtex4-2}

\usepackage{amsmath}
\usepackage{subcaption}
\usepackage{amsfonts}
\usepackage{natbib}
\usepackage{bm}
\usepackage{adjustbox}
\usepackage{multirow}
\usepackage{makecell}
\usepackage{booktabs}
\usepackage{svg}  
\usepackage{tabularx}
\usepackage{cellspace} % 导言区添加
\begin{document}

\title{Noise-Induced Quantum Mpemba Effect}%

\author{Mingrui Zhao, Zhonghuai Hou}%
\email[Email:]{hzhlj@ustc.edu.cn}
\affiliation{Hefei National Laboratory, University of Science and Technology of China, Hefei 230088, China}
\date{March 2025}%
\begin{abstract}
The quantum Mpemba effect (QMPE), an intriguing anomalous relaxation phenomenon, has recently attracted significant attention. However, how various types of noise, which are ubiquitous in real systems, may affect the QMPE remains unknown. Here, we address this gap by constructing a general dynamical framework for $d$ level open quantum systems under random telegraph noise. By investigating the dynamics of an extended system and then projecting back, we find that noise can induce additional modes and strongly influence the relaxation dynamics of the original system. Specially, in the limit of long correlation time of noise, these modes cause
anomalous slowdown for certain initial states, thereby inducing or eliminating QMPE, illustrated by a three-level example system. Interestingly, this mechanism leads to a counter intuitive effect\textemdash the decoherence rate may be slowed down by noise.
\end{abstract}
\maketitle
\textit{Introduction}\textemdash The Mpemba effect denotes a class of counterintuitive anomalous relaxation behaviors, most famously illustrated by the observation that hot water may freeze faster than cold. It was first proposed by Aristotle and rediscovered by Mpemba and Osborne in 1969\cite{Mpemba_1969}. Recently, a fresh understanding of Mpemba effect in general Markovian system was given in\cite{doi:10.1073/pnas.1701264114}, where it was shown that the Mpemba effect occurs when an initial state farther from equilibrium exhibits a smaller projection onto the slow mode. Later, it has been generalized in many classical systems, such as colloidal systems\cite{WOS:000618099100001,WOS:000776483700002}, active particle\cite{WOS:001398165000004}, granular gases\cite{PhysRevLett.119.148001,PhysRevE.99.060901,Biswas_2021,WOS:000677506400004,WOS:000876199300001}, optical resonators\cite{WOS:000424693900003,WOS:000555079800001}, inertial suspensions\cite{PhysRevE.103.032901}, heat engines\cite{,PhysRevE.105.014104}, etc. Moreover, the Mpemba effect has been extended to quantum systems. In the early stages, research on the QMPE mainly focused on open quantum system, where it is usually analyzed using the quantum master equation to investigate how relaxation can be accelerated by carefully choosing the system’s initial state\cite{PhysRevLett.127.060401,PhysRevA.106.012207,PhysRevLett.133.140404,PhysRevE.108.014130}. Within this framework, the thermal Mpemba effect was first discussed in quantum dot systems \cite{PhysRevLett.131.080402} and later confirmed experimentally \cite{PhysRevLett.127.020602,PhysRevLett.133.010403}. Additionally, recent studies have begun to explore the QMPE in isolated quantum many-body systems. Here, concepts such as entanglement asymmetry\cite{WOS:001166947800004} have been used to generalize the Mpemba effect in the context of symmetry restoration\cite{WOS:001144944300001,PhysRevLett.133.140405,INSPEC:25180025,PRXQuantum.6.010324}. On the basis of these findings, theoretical insights have clarified the microscopic origin of this type of QMPE in integrable systems\cite{physrevlett.133.010401}, which have, in turn, motivated experimental realizations in trapped-ion quantum simulators\cite{PhysRevLett.133.010402}. \\
In most studies of the QMPE in open quantum system, the influence of external noise has been largely overlooked. However, external noise is prevalent in practical situations, such as $1/f^{\alpha}$ noise in singlet-triplet qubit\cite{PhysRevLett.110.146804,revmodphys.86.361}, ohmic noise in quantum Hall edge channels\cite{stabler2024giantheatfluxeffect}, colored noise for open quantum battery\cite{10.1063/5.0247924}, random telegraph noise\cite{PhysRevB.94.235433}, to name a few. How these kind of noise may influence QMPE remains unknown. Here, we systematically study the dynamics of $d$ level open quantum system under random telegraph noise (RTN), and analyze its impact on the QMPE. We find that mathematically the time evolution of the system can be understood as a projection of a high-dimensional vector, resulting in additional dynamical modes in the relaxation process. In the long correlation-time limit of RTN, this mechanism leads to anomalous relaxation slowdown for certain initial states, thereby allowing RTN to decisively induce or eliminate the QMPE, depending on the scenario. Furthermore, we show that this effect can even slow down the rate of decoherence, revealing an unexpected dynamical regime that may have implications beyond the QMPE itself.\\
\textit{Setup}\textemdash We consider a d-level system is described by GKSL master equation\cite{thetheoryofopenquanutumsystem}
\begin{equation}
    \frac{d}{dt}\rho=-i[H,\rho]+\Sigma_k\gamma_k(L_k\rho{L_K^{\dag}}-\frac{1}{2}\{L_k^{\dag}L_k,\rho\})
    \label{Lindblad}
\end{equation}where $H=H_0+H_1$, $H_0$ is the original system Hamiltonian and $H_1=\Delta_1\eta\sum_{i,j,i\neq{j}}J_{ij}|i\rangle\langle{j}|$ (with $J_{ij}=J_{ji}^*$ for Hermitian property, $\Delta_1$ describing the strength of noise) denotes the noise perturbation. $\eta$ is the random telegraph noise fulfill $\langle\eta(t)\eta(s)\rangle=exp(-\nu|t-s|)$ and $\langle\eta\rangle=0$. Using Shapiro-Loginov theorem\cite{SHAPIRO1978563}, one can introduce an extended-system whose state can be described by the vector $G=(\langle\rho_{11}\rangle,\langle\rho_{12}\rangle, ..., \langle\eta\rho_{11}\rangle, \langle\eta\rho_{12}\rangle)^T$, of which the dynamics is given by
\begin{equation}
    \frac{d}{dt}G
    =\begin{pmatrix}
        & W_0 & \Delta \\
        & \Delta & W_1
    \end{pmatrix}
    G
    \label{Liouville}
\end{equation}
 $W_0$ is the initial time evolution operator for the original density matrix element$\langle\rho_{ij}\rangle$ ($\langle\cdot\rangle$ stands for the average over telegraph noise), $W_1=W_0-\nu{I}$ is the time evolution operator for $\langle\eta\rho_{ij}\rangle$,and $\Delta=\Delta_1M$ (M is a constant matrix) describes the transfer between $\langle\rho_{ij}\rangle$ and $\langle\eta\rho_{ij}\rangle$. 
 Noting that in the weak-noise limit $\Delta_1\rightarrow0$, by using the non-hermitian perturbation theory\cite{physrevc.6.114,supplement}, we can show that the eigenvalues of $W$ are simply given by $\{\lambda_1=0,\lambda_2,...\lambda_n\}$ and $\{\lambda_1-\nu,\lambda_2-\nu,...\lambda_n-\nu\}$ wherein $\{\lambda_i\}$ are the eigenvalue set of $W_0$ with the order $0>\mathrm{Re}(\lambda_2)>\mathrm{Re}(\lambda_3)>...>\mathrm{Re}(\lambda_n)$, where $n=d^2$. Starting from an initial state $G(0)$, the solution of $G$ can be written as
\begin{align}
    G(t) = &G_{ss} + \sum_{k=2}^{n}e^{\lambda_kt}\langle{L_k|G(0)}\rangle{R_k}+\\
    &\sum_{k=1}^{n}e^{(\lambda_k-\nu)t}\langle{L_{k,-\nu}|G(0)}\rangle{R_{k,-\nu}}
    \nonumber
    \label{dynamicsG}
\end{align}
\begin{table}
    \centering
    \begin{tabular}{ScScSc}
        \toprule
        Picture & Relation of $C^{(i)}_{\mu,-\nu}$ & Effect of noise \\
        \midrule
        % 第一张图片（合并前两行）
        \multirow{2}{*}[16pt]{{\includegraphics[width=1.1in, keepaspectratio]{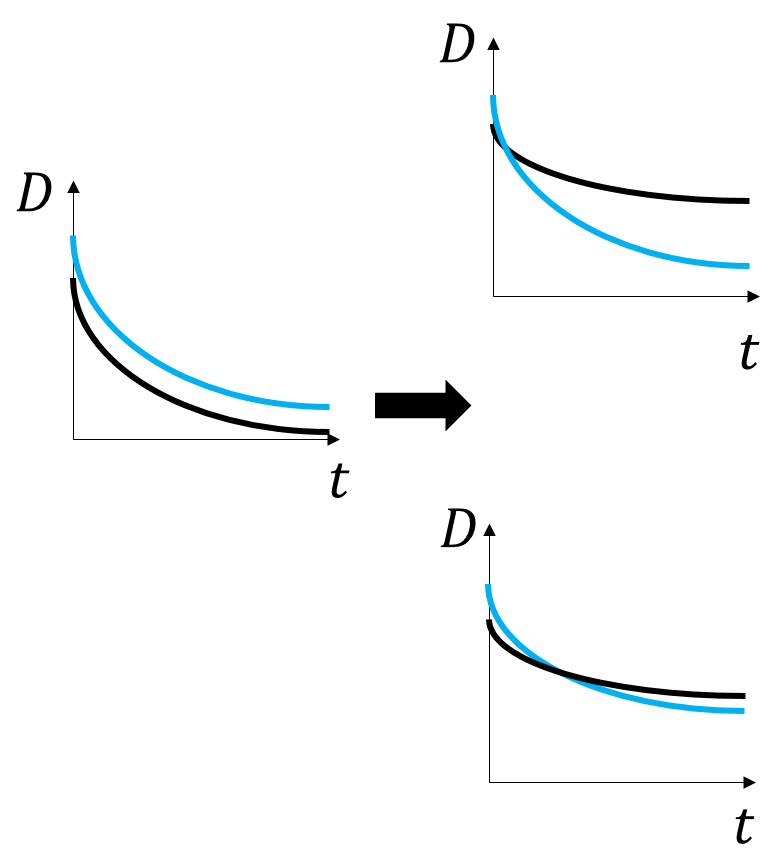}}} & 
        \makecell[c]{$|C^{\alpha}_{\mu,-\nu}|<|C^{\beta}_{\mu,-\nu}|$} & 
        \makecell[c]{Case 1:Induce \\ weak QMPE} \\
        & $C^{\alpha}_{\mu,-\nu}=0,C^{\beta}_{\mu,-\nu}\neq0$ & 
        \makecell[c]{Case 2:Induce \\ strong QMPE} \\
        \midrule
        % 第二张图片（合并后两行）
         \multirow{2}{*}[16pt]{{\includegraphics[width=1.1in, keepaspectratio]{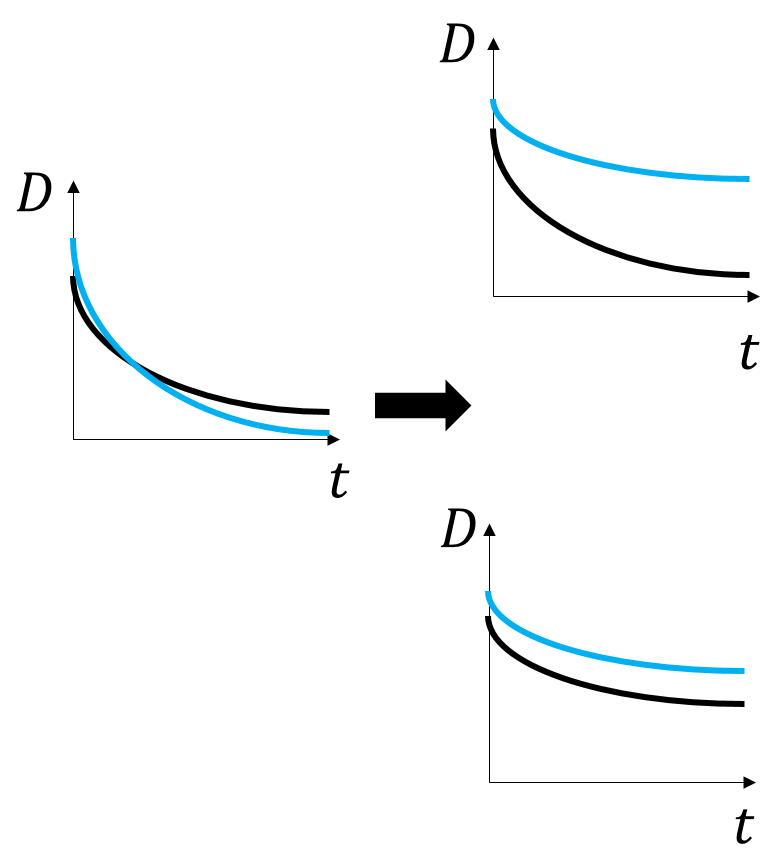}}} & 
        $C^{\alpha}_{\mu,-\nu}\neq0,C^{\beta}_{\mu,-\nu}=0$ & 
        \makecell[c]{Case 3:Eliminate \\ QMPE} \\
        & \makecell[c]{$|C^{\alpha}_{\mu,-\nu}|>|C^{\beta}_{\mu,-\nu}|$} & 
        \makecell[c]{Case 4:Eliminate \\ QMPE} \\
        \bottomrule
    \end{tabular}
    \caption{The blue line stands for the relaxation of initial state $p_{\alpha}(0)$ and the black line stands for $p_{\beta}(0)$. We set the distance to the equilibrium state of $p_{\beta}(0)$ is shorter than $p_{\alpha}(0)$.}
    \label{tab:my_label}
\end{table}
Here $\langle\cdot,\cdot\rangle$ stands for the inner product over the extended system, $G_{ss}$ is the steady state satisfying $dG_{ss}/dt=0$, $(L_k,R_k)$ and $(L_{k,-\nu},R_{k,-\nu})$ are respectively the left and right eigenvectors associated with $\lambda_k$ and $\lambda_{k}-\nu$. \\ 
To get the dynamics of the original system, which one is interested in,we can introduce an operator $\Pi=(\mathcal{I}_{n\times{n}},0_{n\times{n}})$ and apply it to the both side of Eq.(3), where $\mathcal{I}_n\times n$ is the identity operator. After some manipulation\cite{supplement}, we finally obtain the system dynamics described by $\bm{p}=\Pi{G}=(\langle\rho_{11}\rangle,\langle\rho_{12}\rangle,...\langle\rho_{dd}\rangle)^T$
\begin{align}
    \bm{p}(t) = &\bm{p}_{ss} + \sum_{k=2}^{n}e^{\lambda_kt}[l_k|\bm{p}(0)]r_k+\\
    &\sum_{k=1}^{n}e^{(\lambda_k-\nu)t}C_{k,-\nu}\Pi{R_{k,-\nu}}
    \nonumber
    \label{projection1}
\end{align}
where $l_k(r_k)$ are the normalized left(right) eigenvalue of $W_0$ associated with $\lambda_k$, $\bm{p}(0)=\Pi{G}(0)$ and $\bm{p}_{ss}=\Pi{G}_{ss}$ are the initial and steady state of the \textit{original} system, and $[\cdot,\cdot]$ now denotes the inner product over the original system. The coefficient $C_{k,-v}$ is given by
\begin{equation}
    C_{k,-\nu}=\sum_{j\neq{k}}\frac{[l_k|\Delta|r_j]}{\lambda_k-\lambda_j-\nu}\langle{L_j}^{(0)}|G(0)\rangle
\end{equation}
with $L_j^{(0)}=(l_j,0_{n\times{n}})$.\\ Eq.(4) is our main result. The first and second term on the right hand side describe the effect of $W_0$ on the density matrix and is exactly the same as original dynamics without external noise, wherein the slowest relaxation mode is given by $r_2$ with eigenvalue $\lambda_2$ and the coefficient $a_2=[l_2|\bm{p}(0)]$. In the presence of noise, however, the third term matters and new(mixed) mode $\Pi{R_{k,-\nu}}$ with eigenvalues $\lambda_{k,-\nu}$ and coefficient $C_{k,-\nu}$ shows up, which could change substantially the long-time relaxation behavior of $\bm{p}(t)$. Specifially, in the limit of $\nu\rightarrow{0}$, i.e., long-time correlation limit of the noise, the relaxation time scale will be dominated by $|\mathrm{Re}(\lambda_1)-\nu|^{-1}=1/\nu$ rather than $1/|\mathrm{Re}(\lambda_2)|$. More importantly, the introduction of the additional modes may leads to some interesting phenomena, like inducing QMPE or eliminating QMPE, as we will illustrate below. \\ \\
Consider two specific initial states given by 
\begin{equation}
\bm{p}_{\alpha}\left(0\right)=\bm{p}_{ss}+a_{\alpha}r_{\alpha},\text{ }\bm{p}_{\beta}\left(0\right)=\bm{p}_{ss}+a_{\beta}r_{\beta}
\end{equation}
with $\alpha<\beta$ such that $|\mathrm{Re}(\lambda_\alpha)|<|\mathrm{Re}(\lambda_\beta)|$
and $r_{\alpha}$ is the slower mode. In the absence of noise, one
simply has $p_{i}\left(t\right)=\bm{p}_{ss}+a_{i}e^{\lambda_{i}t}r_{i}$
($i=\alpha,\beta$). When the noise is present, however, the time
evolutions are given by, according to Eq.(4), 
\begin{align}
\nonumber
&\bm{p}_{i}\left(t\right)=\bm{p}_{ss}+a_{i}e^{\lambda_{i}t}r_{i}+\\
&C_{\mu,-\nu}^{i}e^{(\lambda_{\mu}-\nu)t}\Pi{R}_{\mu,-\nu}+\text{(faster terms)}\text{ (\ensuremath{i=\alpha,\beta})}
\end{align}
with the coefficients 
\begin{equation}
C_{\mu,-v}^{i}=\frac{a_{i}\left[l_{\mu}|\Delta|r_{i}\right]}{\lambda_{\mu}-\lambda_{i}-\nu},
\end{equation}
where we have used $G_{i}(0)=(\bm{p}_{i}(0),0)$ such that $\left\langle L_{j}^{(0)}|G_{i}(0)\right\rangle =a_{i}\delta_{ij}$,
and $\mu$ denotes the lowest term that survives ($C_{\mu,-\nu}\ne0$)
in the summation over $k$ in Eq.(4). Clearly, if $\nu$ is small enough
such that $\left|\text{Re}\left(\lambda_{\mu}\right)-\nu\right|<\left|\text{Re}\left(\lambda_{\alpha}\right)\right|$,
the slowest mode governing the relaxation dynamics of $\bm{p}_{\alpha,\beta}\left(t\right)$
in the long-time limit would be the mixed-mode $\Pi{R}_{\mu,-\nu}$
with time scale given by $1/\left|\text{Re}\left(\lambda_{\mu}\right)-\nu\right|$.\\ \\
We define the distance of $\bm{p}_{i}$ to system equilibrium state $\bm{p}_{ss}$ as the norm of vector $\bm{p}_{i}-\bm{p}_{ss}$, denoted as $D(\bm{p}_{i}||\bm{p}_{ss})$. If $\left|a_{\alpha}\right|>\left|a_{\beta}\right|$, then no QMPE
exists in the original system since $\bm{p}_{\beta}\left(0\right)$ is
closer to the steady state and decays exponentially faster compared
to $\bm{p}_{\alpha}\left(0\right)$. Now there are two interesting cases
that noise may induce QMPE: \\ \\
1. If both $C_{\mu,-\nu}^{\alpha}$ and $C_{\mu,-\nu}^{\beta}$ are nonzero
and $\left|C_{\mu,-\nu}^{\beta}\right|>\left|C_{\mu,-\nu}^{\alpha}\right|$,
then $D\left[\bm{p}_{\alpha}(t)||\bm{p}_{ss}\right]<D\left[\bm{p}_{\beta}(t)||\bm{p}_{ss}\right]$
if $t$ is large enough. Therefore, the system shows 'weak' QMPE wherein
both states become slow and the initially closer state $\bm{p}_{\beta}(0)$
finally relaxes slower. \\ \\
2. If $C_{\mu,-\nu}^{(\alpha)}=0$ and $C_{\mu,-\nu}^{(\beta)}\ne0$, then $\bm{p}_{\beta}(t)$ is now dominated by a slower mode with eigenvalue
$\lambda_{\mu}-\nu$ while $\bm{p}_{\alpha}(t)$ is dominated by other faster modes. Therefore, the system would show 'strong' QMPE
because $\bm{p}_{\beta}(0)$ now decays exponentially slower than $\bm{p}_{\alpha}(0))$.\\ \\
On the other hand, when $\left|a_{\alpha}\right|<\left|a_{\beta}\right|$
such that the system without noise can show QMPE since now $\bm{p}_{\alpha}(0)$
is closer to $\bm{p}_{ss}$ while $\bm{p}_{\beta}(0)$ decays exponentially
faster. Following similar analyis, one can see that such QMPE may
be eliminated if $\left|C_{\mu,-\nu}^{(\beta)}\right|>\left|C_{\mu,-\nu}^{(\alpha)}\right|\ne0$,
wherein both states decays slower and $\bm{p}_{\alpha}\left(t\right)$
remains closer to $\bm{p}_{ss}$ at large $t$, or if $C_{\mu,-\nu}^{(\alpha)}=0$
and $C_{\mu,-\nu}^{(\beta)}\ne0$ wherein $\bm{p}_{\alpha}(t)$ decays exponentially
faster than $\bm{p}_{\beta}\left(t\right)$. Such interesting effects of
noise induced/eliminated QMPE are outlined in Table 1. 
\\ \\
To a short summary, here we demonstrate a novel type of noise induced
effects associated with QMPE, the underlying mechanism lies in the
attachment of system relaxation dynamics to noise-induced slow modes
if the noise correlation time is large enough. Firstly, this type
of QMPE is totally different to previous ones in the literature where
one mainly focused on exponentially acceleration of dynamics via
tuning the coefficient $a_{2}$\cite{PhysRevLett.127.060401}. The time evolution of $\bm{p}$(Eq.(4)) is the projection of high-dimensional vector so it's actually non-Markovian. Secondly, such noise-induced slowing
down is quite counter-intuitive since generally noise is thought to
accelerate the relaxation dynamics. In particular, this may lead to
a novel phenomenon that noise will decelerate the decoherence of the
system as demonstrated in the example shown below. Finally, we note
that for more general initial states(e.g. two initial states showing weak QMPE in the absence of noise) rather than those specialized
in Eq.(6), similar effects shown in TABLE I also exist.  \\ \\ 
\textit{Examples}\textemdash Consider a three-level quantum optics system described by GKSL master equation. Its dynamics can be described by following equation\cite{thetheoryofopenquanutumsystem}.
\begin{equation}
    \frac{d}{dt}\rho=-i[H_0+H_1,\rho]+D(\rho)
\end{equation}
where we take
\begin{equation}
    H_0=\begin{pmatrix}
        0 & 0 & 0 \\
        0 & \hbar\omega_2 & 0 \\
        0 & 0 & \hbar\omega_3
    \end{pmatrix},
    H_1=\begin{pmatrix}
        0 & 0 & 0 \\
        0 & 0 & \Delta_1\eta \\
        0 & \Delta_1\eta & 0 
    \end{pmatrix}
\end{equation}
We consider the zero-temperature case, i.e. the dissipative part $D(\rho)$ can be written as  
\begin{figure}
\centering
\includegraphics[width=0.48\textwidth]{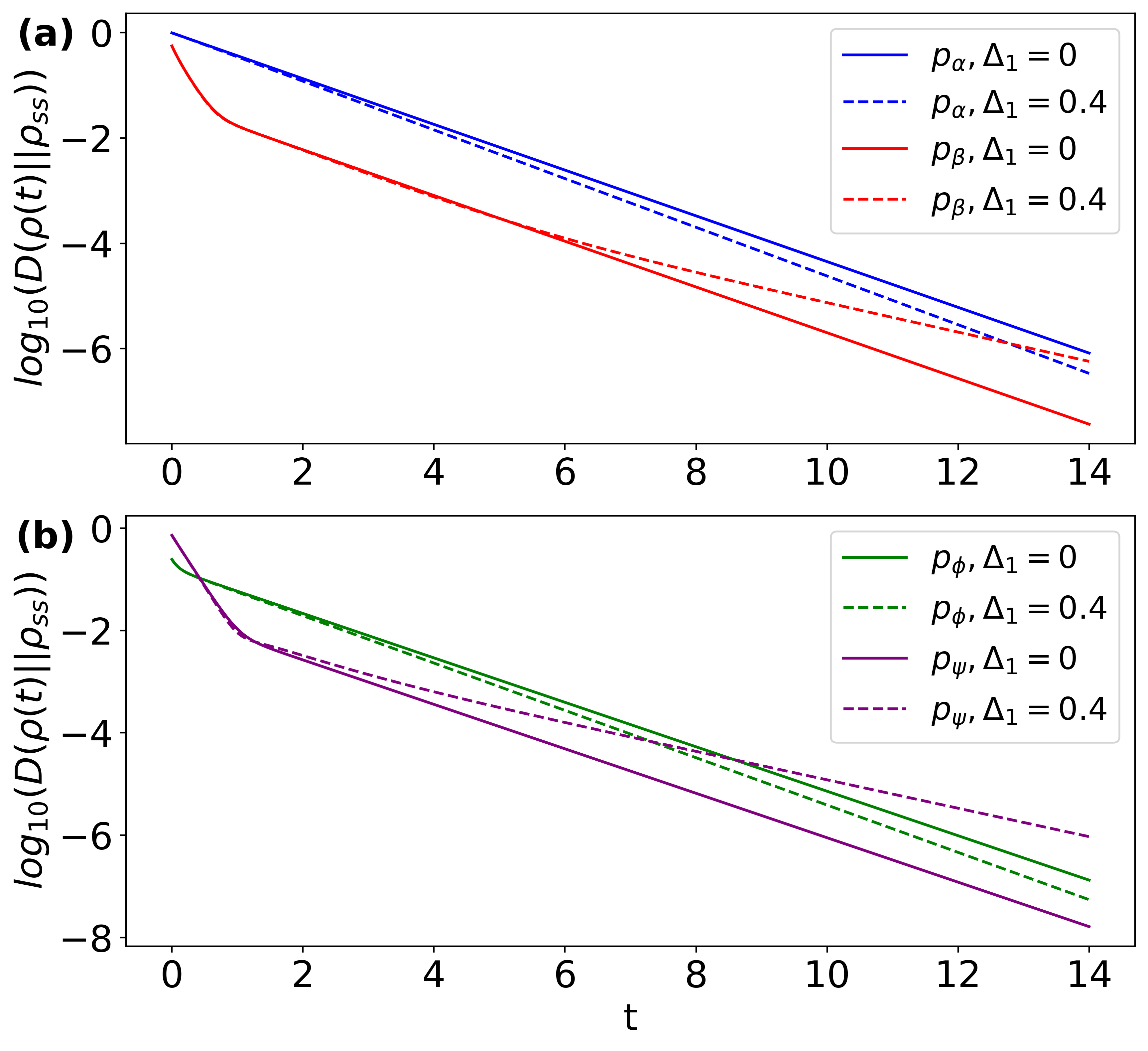}
\captionsetup{font=small, justification=centering}
\caption{The solid lines in the picture depict the time-evolution of given initial states without noise, while the dashed lines depict the one with noise. Noise induces QMPE for $\bm{p}_\alpha,\bm{p}_\beta$((a),dashed line shows crossing while solid line doesn't) and eliminates QMPE for $\bm{p}_\psi,\bm{p}_\phi$((b),dashed line shows double crossing)}
\label{whatcanisay}
\end{figure}
\begin{equation}
    D(\rho)=\sum_{\omega>0}r_{\omega}(L_{\omega}\rho{L^{\dag}_{\omega}}-\frac{1}{2}\{L^{\dag}_{\omega}L_{\omega},\rho\})
\end{equation}
Where $L_{\omega}=|i\rangle\langle{j}|$ with $E_j-E_i=\hbar\omega$.
\begin{table}
\centering
\begin{tabular}{cccc}
\hline
Initial State & $|a_3^{i}|$ & $|C_{2,-\nu}^{i}|$ &  $D(\bm{p}_i(0)||\bm{p}_{ss})$  \\
\hline
$\bm{p}_\alpha$ & $0.69$ & $0$ & $0.99$\\
\hline
$\bm{p}_\beta$ & $0.03$ & $0.04$ & $0.56$\\
\hline
$\bm{p}_\phi$ & $0.11$ & $0$ & $0.24$  \\
\hline
$\bm{p}_\psi$ & $0.01$ & $0.07$ & $0.72$  \\
\hline
\end{tabular}
\caption{Propoties of four different initial states. All data are accurate to two decimal places. We set $\Delta_1=0.4,\nu=0.1,\hbar=1$.Other parameter can be checked in \cite{supplement}.}
\end{table}
Here we choose four different initial states of the system $\bm{p}_\alpha(0),\bm{p}_\beta(0),\bm{p}_\phi(0),\bm{p}_\psi(0)$, and the initial states of the extended system is $(\bm{p}_i(0),0_{n\times1})^T$  since we apply the noise at $t=0$. They both can be written in the form 
\begin{equation}
\bm{p}_i=\bm{p}_{ss}+a_3^{i}r_3+C_{2,-\nu}^{i}\Pi{R}_{2,-\nu}+\mathrm{f.t.}    
\end{equation}
where f.t. is the abbreviate of faster terms. TABLE II lists the properties of these states. The explicit form of these states and the calculation of $C_{2,-\nu}^{i}$ using non-hermitian perturbation theory can be seen in SI\cite{supplement}.  \\ 
First, we consider the states $\bm{p}_{\alpha}$ and $\bm{p}_{\beta}$, corresponding to Case 2 in TABLE I. In the absence of noise, these states do not exhibit the QMPE, as $|a_3^{\beta}| = 0.03 < 0.69 = |a_3^{\alpha}|$ and $D(\bm{p}_{\alpha}(0)||\bm{p}_{ss}) = 0.99 > 0.56 = D(\bm{p}_{\beta}(0)||\bm{p}_{ss})$ (see the solid lines in FIG. 1(a)). When noise is present, $\bm{p}_{\alpha}$ decays slightly faster because the second-order perturbation to the eigenvalue of $W$ is negative. Meanwhile, since $|C_{2,-\nu}^\beta| \neq 0$, $\bm{p}_{\beta}$ exhibits a clear turning point. Before the turning point, $\bm{p}_{\beta}$ decays rapidly along fast mode. After the turning point, the noise-induced slow mode $\lambda_2-\nu$ begin to manifest, so that $\bm{p}_{\beta}$ relaxes more slowly than $\bm{p}_{\alpha}$ in the long-time regime, as shown by the dashed lines in FIG. 1(a). Therefore, this example demonstrates that noise can induce the QMPE.\\ \\
Second, we turn to the states $\bm{p}_{\phi}$ and $\bm{p}_{\psi}$, representing the scenario described in Case 3 of TABLE I. In the absence of noise, these two states exhibit the QMPE, as indicated by the solid lines in FIG. 1(b). The initially farther state, $\bm{p}_{\psi}$, decays faster because $|a_3^{\phi}| = 0.11 > 0.01 = |a_3^{\psi}|$. When noise is introduced, $\bm{p}_{\psi}$ displays a clear turning point and ultimately relaxes more slowly than $\bm{p}_{\phi}$ in the long-time regime, as shown by the dashed lines in FIG. 1(b). This behavior results in a double crossing—constituting a special case of QMPE elimination—since the state $\bm{p}_{\psi}$, which is initially farther from equilibrium, eventually relaxes more slowly in the long-time limit.\\ \\
Interestingly, as mentioned above, the mechanism by which noise induces the QMPE is through the deceleration of the dynamics of certain typical states. Here, we demonstrate that this mechanism can also give rise to a counterintuitive phenomenon---namely, that noise can slow down the rate of decoherence. To illustrate this, we use the coherence relative entropy\cite{RevModPhys.89.041003}, defined as
\begin{equation}
C(\rho(t)) = S(\mathrm{diag}[\rho(t)]) - S(\rho(t))
\end{equation}
where $S(\rho(t))$ is the von Neumann entropy and $\mathrm{diag}[\rho(t)]$ is the diagonal part of $\rho(t)$. The coherence relative entropy serves as a quantitative measure of quantum coherence in a density matrix; its magnitude directly reflects the coherence strength---higher values correspond to stronger quantum coherence, whereas a value of zero indicates a completely incoherent state.\\
We plot the time evolution of the logarithm of $C(\rho_{\beta}(t))$ in FIG. 2 for different values of $\nu$ and $\Delta_1$, where $\rho_{\beta}$ denotes the density matrix corresponding to the initial state $\bm{p}_{\beta}$. Compared to the noiseless case (blue line), the decay of coherence relative entropy exhibits a slowdown in all cases with noise. The red dashed lines illustrate the effect of $\nu$: as $\nu$ decreases, the absolute value of the slope after the turning point becomes smaller, which is consistent with that long-time relaxation is dominated by the mode with eigenvalue $\lambda_2 - \nu$. The purple dot-dashed lines show the effect of noise strength. As the noise intensity increases, $C_{2,-\nu}$ becomes larger [see Eq.~(8)], indicating that the component of $\bm{p}_{\beta}$ along the additional noise-induced mode increases and thus the turning point appears earlier. \\ \\
\textit{Conclusion}\textemdash{We} have demonstrated that external telegraph noise, as a non-Gaussian and non-Markovian perturbation, can induce or eliminate the quantum Mpemba effect (QMPE) in open quantum systems. Unlike the Markovian QMPE driven by initial state engineering \cite{PhysRevLett.127.060401,PhysRevA.106.012207,PhysRevLett.133.140404,PhysRevE.108.014130}, This noise-induced effect originates from the noise-induced additional slow mode, which leads to the relaxation slowdown of specific initial states. Crucially, the noise introduces counter intuitive deceleration of decoherence (FIG. \ref{entropy}), challenging the conventional wisdom that noise invariably accelerates relaxation.
\begin{figure}
\centering
\includegraphics[width=0.48\textwidth]{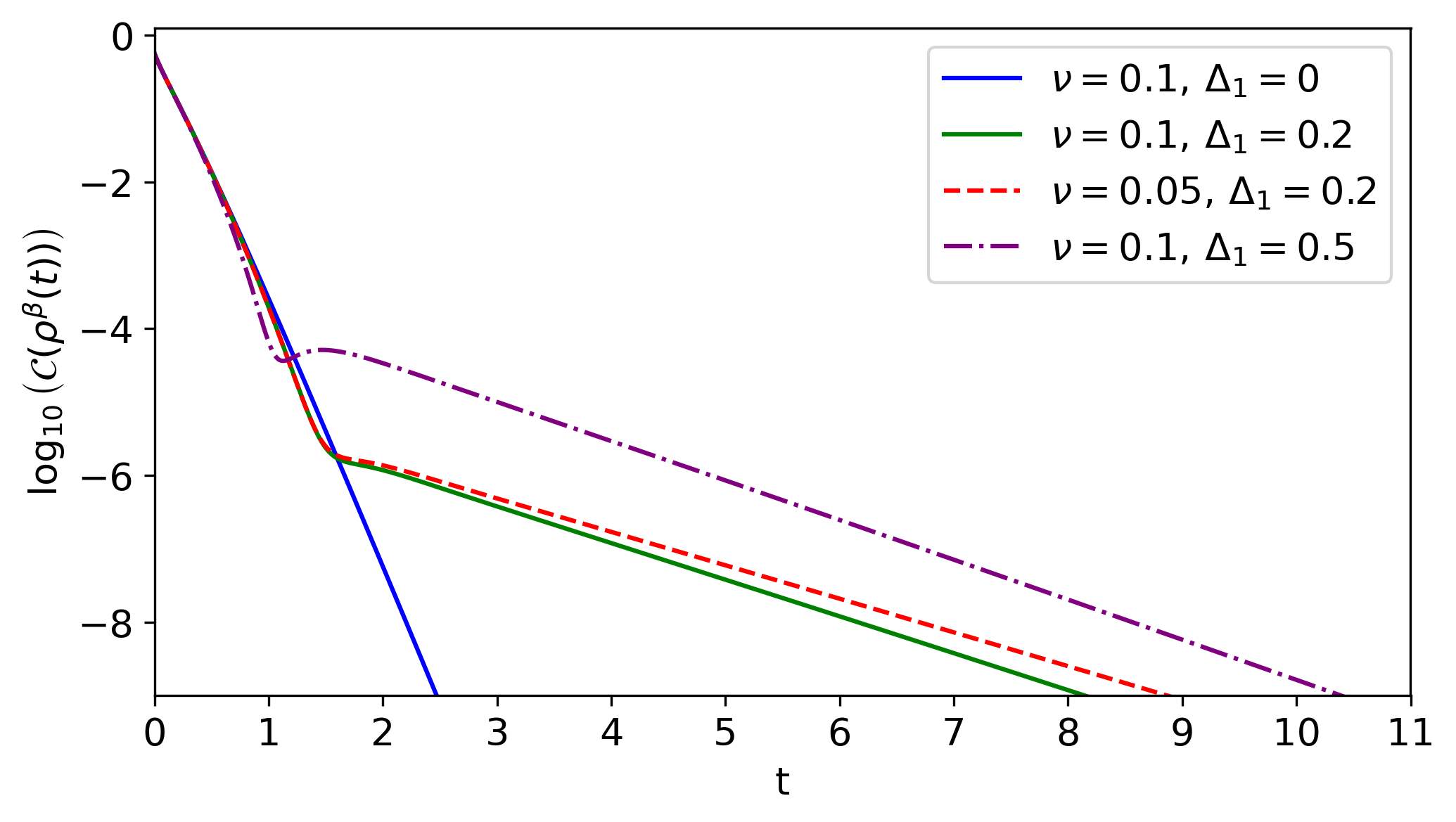}
\captionsetup{font=small, justification=centering}
\caption{This figure illustrates how the logarithm of relative entropy of coherence changes with time.}
\label{entropy}
\end{figure}
This effect stems from the long-time correlation property of noise($\nu$ is very small). When $\nu$ becomes sufficiently large (for example, in the earlier example where $\mathrm{Re}(\lambda_2)-\nu=\lambda_3$), this phenomenon disappears. In the limit of $\nu\rightarrow+\infty$, RTN behaves similarly to white noise, introducing an additional dissipative term into the original Lindblad equation\cite{supplement}.\\ 
Our work opens two practical avenues. First for quantum control: Tailored noise could selectively protect coherence in specific subsystems (e.g. quantum memory), as evidenced by the delayed decoherence in FIG. \ref{entropy}.
Secend, telegraph noise might enhance parameter estimation in critical systems where QMPE amplifies sensitivity \cite{PhysRevLett.131.050801}. Future studies could explore optimal noise spectra for these applications, or generalize our framework to other non-Gaussian noises, like non-Gaussian ohm noise.\\ \\
This work is supported by Innovation Program for Quantum Science and Technology(2021ZD0303300).
\bibliographystyle{unsrt}
\bibliography{references}

\begin{thebibliography}{10}

\bibitem{Mpemba_1969}
E~B Mpemba and D~G Osborne.
\newblock Cool?
\newblock {\em Physics Education}, 4(3):172, may 1969.

\bibitem{doi:10.1073/pnas.1701264114}
Zhiyue Lu and Oren Raz.
\newblock Nonequilibrium thermodynamics of the markovian mpemba effect and its
  inverse.
\newblock {\em Proceedings of the National Academy of Sciences},
  114(20):5083--5088, 2017.

\bibitem{WOS:000618099100001}
Avinash Kumar and John Bechhoefer.
\newblock Exponentially faster cooling in a colloidal system.
\newblock {\em NATURE}, 584(7819):64+, AUG 6 2020.

\bibitem{WOS:000776483700002}
Avinash Kumar, Raphael Chetrite, and John Bechhoefer.
\newblock Anomalous heating in a colloidal system.
\newblock {\em PROCEEDINGS OF THE NATIONAL ACADEMY OF SCIENCES OF THE UNITED
  STATES OF AMERICA}, 119(5), FEB 1 2022.

\bibitem{WOS:001398165000004}
Apurba Biswas and R.~Rajesh.
\newblock Mpemba effect in the relaxation of an active brownian particle in a
  trap without metastable states.
\newblock {\em JOURNAL OF CHEMICAL PHYSICS}, 162(3), JAN 21 2025.

\bibitem{PhysRevLett.119.148001}
Antonio Lasanta, Francisco Vega~Reyes, Antonio Prados, and Andr\'es Santos.
\newblock When the hotter cools more quickly: Mpemba effect in granular fluids.
\newblock {\em Phys. Rev. Lett.}, 119:148001, Oct 2017.

\bibitem{PhysRevE.99.060901}
Aurora Torrente, Miguel~A. L\'opez-Casta\~no, Antonio Lasanta, Francisco~Vega
  Reyes, Antonio Prados, and Andr\'es Santos.
\newblock Large mpemba-like effect in a gas of inelastic rough hard spheres.
\newblock {\em Phys. Rev. E}, 99:060901, Jun 2019.

\bibitem{Biswas_2021}
Apurba Biswas, V.~V. Prasad, and R.~Rajesh.
\newblock Mpemba effect in an anisotropically driven granular gas.
\newblock {\em Europhysics Letters}, 136(4):46001, mar 2022.

\bibitem{WOS:000677506400004}
E.~Mompo, M.~A. Lopez-Castano, A.~Lasanta, F.~Vega~Reyes, and A.~Torrente.
\newblock Memory effects in a gas of viscoelastic particles.
\newblock {\em PHYSICS OF FLUIDS}, 33(6), JUN 2021.

\bibitem{WOS:000876199300001}
Alberto Megias and Andres Santos.
\newblock Mpemba-like effect protocol for granular gases of inelastic and rough
  hard disks.
\newblock {\em FRONTIERS IN PHYSICS}, 10, OCT 6 2022.

\bibitem{WOS:000424693900003}
Tim Keller, Valentin Torggler, Simon~B. Jaeger, Stefan Schuetz, Helmut Ritsch,
  and Giovanna Morigi.
\newblock Quenches across the self-organization transition in multimode
  cavities.
\newblock {\em NEW JOURNAL OF PHYSICS}, 20, FEB 7 2018.

\bibitem{WOS:000555079800001}
Andres Santos and Antonio Prados.
\newblock Mpemba effect in molecular gases under nonlinear drag.
\newblock {\em PHYSICS OF FLUIDS}, 32(7), JUL 1 2020.

\bibitem{PhysRevE.103.032901}
Satoshi Takada, Hisao Hayakawa, and Andr\'es Santos.
\newblock Mpemba effect in inertial suspensions.
\newblock {\em Phys. Rev. E}, 103:032901, Mar 2021.

\bibitem{PhysRevE.105.014104}
Jie Lin, Kai Li, Jizhou He, Jie Ren, and Jianhui Wang.
\newblock Power statistics of otto heat engines with the mpemba effect.
\newblock {\em Phys. Rev. E}, 105:014104, Jan 2022.

\bibitem{PhysRevLett.127.060401}
Federico Carollo, Antonio Lasanta, and Igor Lesanovsky.
\newblock Exponentially accelerated approach to stationarity in markovian open
  quantum systems through the mpemba effect.
\newblock {\em Phys. Rev. Lett.}, 127:060401, Aug 2021.

\bibitem{PhysRevA.106.012207}
Simon Kochsiek, Federico Carollo, and Igor Lesanovsky.
\newblock Accelerating the approach of dissipative quantum spin systems towards
  stationarity through global spin rotations.
\newblock {\em Phys. Rev. A}, 106:012207, Jul 2022.

\bibitem{PhysRevLett.133.140404}
Mattia Moroder, Ois\'{\i}n Culhane, Krissia Zawadzki, and John Goold.
\newblock Thermodynamics of the quantum mpemba effect.
\newblock {\em Phys. Rev. Lett.}, 133:140404, Oct 2024.

\bibitem{PhysRevE.108.014130}
Felix Ivander, Nicholas Anto-Sztrikacs, and Dvira Segal.
\newblock Hyperacceleration of quantum thermalization dynamics by bypassing
  long-lived coherences: An analytical treatment.
\newblock {\em Phys. Rev. E}, 108:014130, Jul 2023.

\bibitem{PhysRevLett.131.080402}
Amit~Kumar Chatterjee, Satoshi Takada, and Hisao Hayakawa.
\newblock Quantum mpemba effect in a quantum dot with reservoirs.
\newblock {\em Phys. Rev. Lett.}, 131:080402, Aug 2023.

\bibitem{PhysRevLett.127.020602}
Fusheng Chen, Zheng-Hang Sun, Ming Gong, Qingling Zhu, Yu-Ran Zhang, Yulin Wu,
  Yangsen Ye, Chen Zha, Shaowei Li, Shaojun Guo, Haoran Qian, He-Liang Huang,
  Jiale Yu, Hui Deng, Hao Rong, Jin Lin, Yu~Xu, Lihua Sun, Cheng Guo, Na~Li,
  Futian Liang, Cheng-Zhi Peng, Heng Fan, Xiaobo Zhu, and Jian-Wei Pan.
\newblock Observation of strong and weak thermalization in a superconducting
  quantum processor.
\newblock {\em Phys. Rev. Lett.}, 127:020602, Jul 2021.

\bibitem{PhysRevLett.133.010403}
Shahaf Aharony~Shapira, Yotam Shapira, Jovan Markov, Gianluca Teza, Nitzan
  Akerman, Oren Raz, and Roee Ozeri.
\newblock Inverse mpemba effect demonstrated on a single trapped ion qubit.
\newblock {\em Phys. Rev. Lett.}, 133:010403, Jul 2024.

\bibitem{WOS:001166947800004}
Filiberto Ares, Sara Murciano, and Pasquale Calabrese.
\newblock Entanglement asymmetry as a probe of symmetry breaking.
\newblock {\em NATURE COMMUNICATIONS}, 14(1), APR 11 2023.

\bibitem{WOS:001144944300001}
Sara Murciano, Filiberto Ares, Israel Klich, and Pasquale Calabrese.
\newblock Entanglement asymmetry and quantum mpemba effect in the xy spin
  chain.
\newblock {\em JOURNAL OF STATISTICAL MECHANICS-THEORY AND EXPERIMENT},
  2024(1), JAN 1 2024.

\bibitem{PhysRevLett.133.140405}
Shuo Liu, Hao-Kai Zhang, Shuai Yin, and Shi-Xin Zhang.
\newblock Symmetry restoration and quantum mpemba effect in symmetric random
  circuits.
\newblock {\em Phys. Rev. Lett.}, 133:140405, Oct 2024.

\bibitem{INSPEC:25180025}
Katja Klobas, Colin Rylands, and Bruno Bertini.
\newblock Translation symmetry restoration under random unitary dynamics.
\newblock {\em Phys. Rev. B}, 111:L140304, Apr 2025.

\bibitem{PRXQuantum.6.010324}
Alessandro Foligno, Pasquale Calabrese, and Bruno Bertini.
\newblock Nonequilibrium dynamics of charged dual-unitary circuits.
\newblock {\em PRX Quantum}, 6:010324, Feb 2025.

\bibitem{physrevlett.133.010401}
Colin Rylands, Katja Klobas, Filiberto Ares, Pasquale Calabrese, Sara Murciano,
  and Bruno Bertini.
\newblock Microscopic origin of the quantum mpemba effect in integrable
  systems.
\newblock {\em Phys. Rev. Lett.}, 133:010401, Jul 2024.

\bibitem{PhysRevLett.133.010402}
Lata~Kh. Joshi, Johannes Franke, Aniket Rath, Filiberto Ares, Sara Murciano,
  Florian Kranzl, Rainer Blatt, Peter Zoller, Beno\^{\i}t Vermersch, Pasquale
  Calabrese, Christian~F. Roos, and Manoj~K. Joshi.
\newblock Observing the quantum mpemba effect in quantum simulations.
\newblock {\em Phys. Rev. Lett.}, 133:010402, Jul 2024.

\bibitem{PhysRevLett.110.146804}
O.~E. Dial, M.~D. Shulman, S.~P. Harvey, H.~Bluhm, V.~Umansky, and A.~Yacoby.
\newblock Charge noise spectroscopy using coherent exchange oscillations in a
  singlet-triplet qubit.
\newblock {\em Phys. Rev. Lett.}, 110:146804, Apr 2013.

\bibitem{revmodphys.86.361}
E.~Paladino, Y.~M. Galperin, G.~Falci, and B.~L. Altshuler.
\newblock 1/f noise: Implications for solid-state quantum information.
\newblock {\em Rev. Mod. Phys.}, 86:361--418, Apr 2014.

\bibitem{stabler2024giantheatfluxeffect}
Florian Stäbler, Alioune Gadiaga, and Eugene~V. Sukhorukov.
\newblock Giant heat flux effect in non-chiral transmission lines.
\newblock 2024.

\bibitem{10.1063/5.0247924}
Zohreh Khodadad, Jalal Tarabishi, and Gabriel Hanna.
\newblock Impact of disorder on the exciton dynamics of an open quantum
  battery.
\newblock {\em The Journal of Chemical Physics}, 162(7):074309, 02 2025.

\bibitem{PhysRevB.94.235433}
L.~K. Castelano, F.~F. Fanchini, and K.~Berrada.
\newblock Open quantum system description of singlet-triplet qubits in quantum
  dots.
\newblock {\em Phys. Rev. B}, 94:235433, Dec 2016.

\bibitem{thetheoryofopenquanutumsystem}
Heinz-Peter Breuer and Francesco Petruccione.
\newblock {\em The Theory of Open Quantum Systems}.
\newblock Oxford University Press, 01 2007.

\bibitem{SHAPIRO1978563}
V.E. Shapiro and V.M. Loginov.
\newblock “formulae of differentiation” and their use for solving
  stochastic equations.
\newblock {\em Physica A: Statistical Mechanics and its Applications},
  91(3):563--574, 1978.

\bibitem{physrevc.6.114}
Morton~M. Sternheim and James~F. Walker.
\newblock Non-hermitian hamiltonians, decaying states, and perturbation theory.
\newblock {\em Phys. Rev. C}, 6:114--121, Jul 1972.

\bibitem{supplement}
See supplemental material for additional derivations, data, and figures.

\bibitem{RevModPhys.89.041003}
Alexander Streltsov, Gerardo Adesso, and Martin~B. Plenio.
\newblock Colloquium: Quantum coherence as a resource.
\newblock {\em Rev. Mod. Phys.}, 89:041003, Oct 2017.

\bibitem{PhysRevLett.131.050801}
Si-Yuan Bai and Jun-Hong An.
\newblock Floquet engineering to overcome no-go theorem of noisy quantum
  metrology.
\newblock {\em Phys. Rev. Lett.}, 131:050801, Aug 2023.

\end{thebibliography}
\clearpage
\appendix
\onecolumngrid % 临时切换回单栏
\section*{Supplemental Material}
\renewcommand{\theequation}{S\arabic{equation}} % Number eq as S1, S2...
\setcounter{equation}{0} % reset eq number

\subsection{The derivation of Eq.(2)}
Here we explain the derivation of equation(2). Using Shapiro-Loginov theorem\cite{SHAPIRO1978563}
\begin{equation}
    \frac{d}{dt}\langle\eta{x}\rangle=\langle\eta\frac{d}{dt}x\rangle-\nu\langle\eta{x}\rangle
\label{theorem}
\end{equation}
Where $\eta$ is the telegraph noise and $\langle\cdot\rangle$ represent the average over the telegraph noise. We consider the dynamical equation (1) in the main letter. Rewrite the density matrix element $\rho_{ij}$ as $\bm{p}$, we get
\begin{align}
    \frac{d}{dt}\rho=\mathcal{L}_0\rho&=-i[H_0,\rho]+\sum_k(L_k\rho{L_k^\dag}-\frac{1}{2}\{L_k^\dag{L}_k,\rho\})\rightarrow\frac{d}{dt}\bm{p}_i=W_{ij}\bm{p}_j \nonumber \\
    &\frac{d}{dt}\rho=\mathcal{L}_1\rho=-i[H_1,\rho]\rightarrow\frac{d}{dt}\bm{p}_i=\Delta_1M_{ij}\eta\bm{p}_j
\end{align}
Note that $\mathcal{L}_0,\mathcal{L}_1$ are both linear maps, so if we rewrite density matrix as vector form $\bm{p}$, these equations can be written in form of matrix $W_{ij}$ and $\Delta_1M_{ij}\eta$. Since (1) is the sum of these two term, we get the total equation for $\bm{p}$:
\begin{equation}
    \frac{d}{dt}\langle\bm{p}_i\rangle=W_{ij}\langle\bm{p}_j\rangle+\Delta_1M_{ij}\langle\eta\bm{p}_j\rangle
    \label{master}
\end{equation}
Consider the time evolution for $\langle\eta\vec{\rho}_i\rangle$ with (\ref{theorem})
\begin{equation}
    \frac{d}{dt}\langle\eta\bm{p}_i\rangle=\langle\eta({W_{ij}}\bm{p}_j+\Delta_1M_{ij}\eta\bm{p}_j)\rangle-\nu\langle\eta\bm{p}_j\rangle
\end{equation}
Using $\eta^2=1$, we get
\begin{equation}
    \frac{d}{dt}\langle\eta\bm{p}_i\rangle=(W_{ij}-\nu)\langle\eta\bm{p}_j\rangle+\Delta_1M_{ij}\langle\bm{p}_j\rangle
    \label{master2}
\end{equation}
Note that $\langle\bm{p}_i\rangle$ and $\langle\eta\bm{p}_i\rangle$ are the element of $G_i$ for $i\in\{1,...d^2\}$ and $i\in \{d^2+1,...2d^2\}$ respectively. Thus we get the expression (2) in the main text.\\ \\
Consider Eq.(S3). Generally the existance of $\Delta_1M_{ij}$ will change the equilibrium state. However, in the small noise limit, noise does not change the relative distances from the two given initial states to the equilibrium state. Interestingly, if the steady state in no-noise case fulfill
\begin{equation}
    M\bm{p}_{ss}=0
\end{equation}
Then the applying of the noise will not change steady state. We will see this fact in our calculation of three-level system. 
\subsection{The derivation of Eq.(4)}
We separate $W$ into two parts:
\begin{equation}
    W=\begin{pmatrix}
        W_0 & 0 \\
        0 & W_1
    \end{pmatrix}+
    \begin{pmatrix}
        0 & \Delta \\
        \Delta & 0
    \end{pmatrix}
\end{equation}
The second part can be seen as the perturbation to the first part. Since the diagonal elements of the second part is zero, the first-order perturbation to the eigenvalue is zero. Consider Eq.(3) in the main text. We set the noise to be applied at $t=0$, so $G(0)=(\bm{p}(0),0_{n\times1})^T$. Since the first part is block-diagonal, it's easy to find that the left and right eigenvalue for $\lambda_k$ of $W$ is $L_k=(l_k,0_{n\times1}),R_k=(r_k,0_{n\times1})^T$, for $\lambda_k-\nu$ is $L_{k,-\nu}=(0_{n\times1},l_k),R_{k.-\nu}=(0_{n\times1},r_k)^T$. Therefore, when $\Delta=0$(no noise case), it's easy to find
\begin{equation}
    \bm{p}(t)=\Pi{G}(t)=\bm{p}_{ss}+\sum_{k=2}^ne^{\lambda_kt}[l_k|\bm{p(0)}]r_k
\end{equation}
which is exactly the same form as the general spectral decomposition form in \cite{PhysRevLett.127.060401}. As the noise is applied, using equations in \cite{physrevc.6.114}, we find the first-order perturbation for $L_k,R_k.L_{k,-\nu},R_{k,-\nu}$ is 
\begin{align}
    L_k^{(1)}=\sum_{n\neq{k}}\frac{[l_k|\Delta|r_n]}{\lambda_k-\lambda_n+\nu}(0_{n\times1},l_n),& R_k^{(1)}=\sum_{n\neq{k}}\frac{[l_n|\Delta|r_k]}{\lambda_k-\lambda_n+\nu}(0_{n\times1},r_n)^T \nonumber \\
    L_{k,-\nu}^{(1)}=\sum_{n\neq{k}}\frac{[l_k|\Delta|r_n]}{\lambda_k-\lambda_n-\nu}(l_n,0_{n\times1}),&R_{k,-\nu}^{(1)}=\sum_{n\neq{k}}\frac{[l_n|\Delta|r_k]}{\lambda_k-\lambda_n-\nu}(r_n,0_{n\times1})^T
\end{align}
These results are very interesting: for $\Delta=0$ case, $\langle L_{k,-\nu}|G(0)\rangle=(0_{n\times1},l_k)(r_k,0_{n\times1})^T=0$, and $\Pi R_{k,-\nu}=(\mathcal{I}_{n\times n},0_{n\times n})(0_{n\times1},r_k)=0$. This fact shows that the extra mode $\lambda_k-\nu$ cannot influence system dynamics, which is in agreement with our intuition. However, as we apply the external noise, $\langle L_{k,-\nu}^{(1)}|G(0)\rangle\neq0$ and $\Pi R_{k,-\nu}^{(1)}\neq0$, indicating that the extra modes induced by the noise begin to affect the system dynamics. \\
For the sake of rigor, we consider the normalization factor after the noise is applied. For mode $\lambda_k$, the normalization factor $b_k$ fulfill
\begin{equation}
    \frac{1}{b_k}=\langle (L_k^{(0)}+L_k^{(1)}|(R_k^{(0)}+R_k^{(1)})\rangle=1+\langle L_k^{(1)}|R_k^{(1)}\rangle
\end{equation}
Note that both $L_k^{(1)},R_k^{(1)}$ are of the order $\Delta_1$, therefore $b_k\approx1-\Delta_1^2$. So in the frame of first order perturbation, we can neglect $b_k$. \\
Putting Eq.(S9) into Eq.(3) in the main text, we finally get Eq.(4). In principle there should be $b_k$ in front of the second term of Eq.(4), but we ignore it for the reason above. 
\subsection{Calculation of three-level system}
Here we give the explicit expression of the operator $W_0,\Delta$ in the main text of the three-level system. 
Here we have 
\begin{equation}
    \bm{p}=(\langle\rho_{11}\rangle,\langle\rho_{22}\rangle,\langle\rho_{33}\rangle,\langle\rho_{12}\rangle,\langle\rho_{21}\rangle,\langle\rho_{13}\rangle,\langle\rho_{31}\rangle,\langle\rho_{23}\rangle,\langle\rho_{32}\rangle)^T
\end{equation}
Also,
\begin{equation}
    W_0=\begin{pmatrix}
        K_1 & 0 \\
        0 & K_2
    \end{pmatrix}
\end{equation}
with 
\begin{equation}
    K_1=\begin{pmatrix}
        0 & \gamma_{21} & \gamma_{31} \\
        0 & -\gamma_{21} & \gamma_{32} \\
        0 & 0 & -\gamma_{32}-\gamma_{21}
    \end{pmatrix}
\end{equation}
and 
\begin{align}
    K_2=\mathrm{diag}(-\frac{1}{2}\gamma_{21}+i\omega_2,-\frac{1}{2}\gamma_{21}-i\omega_2,-\frac{1}{2}(\gamma_{31}+\gamma_{32})+i\omega_3,-\frac{1}{2}(\gamma_{31}+\gamma_{32})-i\omega_3,  \nonumber \\ 
    -\frac{1}{2}(\gamma_{31}+\gamma_{32})-i(\omega_2-\omega_3),-\frac{1}{2}(\gamma_{31}+\gamma_{32})-i(\omega_3-\omega_2))
\end{align}
The explicit form of $\Delta$ is
\begin{equation}
    \Delta=i\Delta_1\begin{pmatrix}
        0 & 0 & 0 & 0 & 0 & 0 & 0 & 0 & 0\\
        0 & 0 & 0 & 0 & 0 & 0 & 0 & 1 & -1 \\
        0 & 0 & 0 & 0 & 0 & 0 & 0 & -1 & 1 \\
        0 & 0 & 0 & 0 & 0 & 1 & 0 & 0 & 0 \\
        0 & 0 & 0 & 0 & 0 & 0 & -1 & 0 & 0 \\
        0 & 0 & 0 & 1 & 0 & 0 & 0 & 0 & 0 \\
        0 & 0 & 0 & 0 & -1 & 0 & 0 & 0 & 0 \\
        0 & 1 & -1 & 0 & 0 & 0 & 0 & 0 & 0\\
        0 & -1 & 1 & 0 & 0 & 0 & 0 & 0 & 0
    \end{pmatrix}
\end{equation}

We set $\gamma_{21}=\gamma_{32}=1,\gamma_{31}=8,\omega_2=1,\omega_3=2$. These expressions can be easily obtained by carefully calculating each term in Eq.9 in the main text. Interestingly, we find in zero-temperature case, the steady state of the system is $\bm{p}_{ss}=(1,0,0,0,0,0,0,0,0)^T$ which fulfill $\Delta\bm{p}_{ss}=0$, therefore the applying of external noise will not change the steady state of the system. \\
Now we give the full expression of initial states $\bm{p}_\alpha,\bm{p}_\beta,\bm{p}_\phi,\bm{p}_\psi$. The corresponding density matrix is
\begin{align}
\nonumber
    &\rho^{\alpha}= \begin{pmatrix}
        0.3 & 0 & 0 \\
        0 & 0.7 & 0 \\
        0 & 0 & 0
    \end{pmatrix}, 
    \rho^{\beta} = \begin{pmatrix}
        0.75 & 0 & \sqrt{3}/4 \\
        0 & 0 & 0 \\
        \sqrt{3}/4 & 0 & 0.25
    \end{pmatrix} \\
    &\rho^{\phi}= \begin{pmatrix}
        0.8 & 0 & 0 \\
        0 & 0.1 & 0 \\
        0 & 0 & 0.1
    \end{pmatrix},
    \rho^{\psi} = \begin{pmatrix}
        8/9 & 0 & 2\sqrt{3}/9 \\
        0 & 0 & 0 \\
        2\sqrt{3}/9 & 0 & 1/9
    \end{pmatrix}
    \label{initialstate}
\end{align}
We consider the calculation using non-hermitian perturbation theory. Since $C_{\mu,-\nu}^i=\langle{L}_{\mu,-\nu}|G_i(0)\rangle$ with $G_i(0)=(\bm{p}_i(0),0_{9\times1})^T$, and $L_{\mu,-\nu}^{(0)}=(0_{9\times1},l_{\mu})$, only first order correction of $L_{\mu,-\nu}$ contributes to $C_{\mu,\nu}^i$.  $L_{\mu,-\nu}$ is the conjugate transpose of the eigenvector of $W^\dag$ with eigenvalue  $\lambda_\mu*$. First we consider $L_{1,-\nu}^{(1)}$. The eigenvector of $W^\dag$ with eigenvalue $\lambda_1-\nu=-\nu$ is $\frac{1}{\sqrt{3}}(0_{9\times1},1,1,1,0,0,0,0,0,0)^T$. Using equation in \cite{physrevc.6.114}, it's easy to find $L^{(1)}_{1,-\nu}=0$, such that $C_{1,-\nu}$ for all initial states are zero. \\ \\
Next we consider $C_{2,-\nu}^i$. We get 
\begin{equation}
    L_{2,-\nu}^{(1)} =\frac{-i\Delta_1}{\frac{1}{2}(\gamma_{31}+\gamma_{32}-\gamma_{21}-\nu)+i(\omega_3-\omega_2)}(0,0,0,0,0,1,0,0,0,0_{9\times1})^T
\end{equation}
Using this it's easy to find that $C_{2,-\nu}^\alpha,C_{2,-\nu}^\phi=0$ and $C_{2,-\nu}^\beta,C_{2,-\nu}^\psi\neq0$. By substituting these parameter values into the expression and performing normalization, one obtains the values reported in the main text. \\
Next we explain the slight acceleration of blue dashed line and green dashed line compare to solid line in the main text using the second order perturbation theory. Using Eq.(4.9) in \cite{physrevc.6.114}, we get 
\begin{equation}
    \lambda_3^{(2)}=\frac{-\Delta_1^2}{\frac{1}{2}(\gamma_{31}+\gamma_{32}-\gamma_{21}+2\nu)+i(\omega_2-\omega_3)}
\end{equation}
It's easy to find that the real part of $\lambda_3^{(2)}$ is minus. Therefore, the second order effect of noise will slightly accelerate relaxation process, as shown in dashed line in Fig.1.
\subsection{RTN in short correlation time limit}
In short correlation time limit($\nu\rightarrow\infty$), RTN has similar effect as Gaussian white noise, and induce extra dissipative term in Lindblad equation. Using the expression in (S2), we consider the form of $\mathcal{L}_1$ in this limit. Using second-order time-dependent perturbation theory, we have
\begin{equation}
    \frac{d}{dt}=\mathcal{L}_1\rho=\Delta_1\eta\mathcal{A}(\rho)
\end{equation}
such that
\begin{align}
\rho(t)=&\rho(0)+\Delta_1\int_0^t\eta(s)\mathcal{A}(\rho(s))ds = \nonumber \\
&\rho(0)+\Delta_1\int_0^t\eta(s)[\rho(0)+\Delta_1\int_0^s\eta(\tau)\mathcal{A}\circ\mathcal{A}(\rho(\tau))d\tau]
\end{align}
Using the property of RTN($\langle\eta(t)\rangle=0,\langle\eta(t)\eta(s)\rangle=exp(-\nu|t-s|)$) and consider above equation under the average over RTN, we get
\begin{equation}
    \langle\rho(t)\rangle=\langle\rho(0)\rangle+\Delta_1^2\int_0^t\int_0^s{exp}(-\nu(s-\tau))\mathcal{A}\circ\mathcal{A}(\rho(\tau))d\tau{d}s
\end{equation}
In the limit of $\nu\rightarrow+\infty$, using the Markovian approximation
\begin{equation}
    \int_0^sexp(-\nu(s-\tau))f(\tau)d\tau=\frac{1}{\nu}f(s)
\end{equation}
we get
\begin{equation}        \langle\rho(t)\rangle=\langle\rho(0)\rangle+\frac{\Delta_1^2}{\nu}\int_0^t\mathcal{A}\circ\mathcal{A}\langle\rho(s)\rangle{ds}
\end{equation}
Thus, we can get 
\begin{equation}
    \frac{d}{dt}\langle\rho(t)\rangle=\frac{\Delta_1^2}{\nu}\mathcal{A}\circ\mathcal{A}\langle\rho(t)\rangle
\end{equation}
In the main text, $\mathcal{L}_1\rho=-i[H_1,\rho]=-i\Delta_1\eta[J,\rho]$. Therefore, in this limit the RTN reduce to an extra dissipative term, and Eq.(1) becomes
\begin{equation}
    \frac{d}{dt}\rho=-i[H_0,\rho]+\Sigma_k\gamma_k(L_k\rho{L_K^{\dag}}-\frac{1}{2}\{L_k^{\dag}L_k,\rho\})
    -[J,[J,\rho]]
\end{equation}
%\bibliographystyle{unsrt}
%\bibliography{references}
\end{document}